\documentclass[journal]{vgtc}                     

\onlineid{3305}

\vgtccategory{Research}

\vgtcpapertype{please specify}

\title{RAGExplorer: A Visual Analytics System for the Comparative Diagnosis of RAG Systems}

\author{%
  Haoyu Tian$^{*}$, 
  \authororcid{Yingchaojie Feng$^{*\dag}$}{0000-0002-1418-4635}, 
  Zhen Wen,
  Haoxuan Li, 
  \authororcid{Minfeng Zhu$^{\dag}$}{0000-0002-6711-3099},
  and Wei Chen$^{\dag}$
}

\authorfooter{
  \item
  	Haoyu Tian, Zhen Wen, Haoxuan Li, and Wei Chen are with the State Key Lab of CAD\&CG, Zhejiang University. 
  	E-mail: \{thyme | wenzhen | lihaoxuan | chenvis\}@zju.edu.cn.
  \item
  	Yingchaojie Feng is with School of Computing, National University of Singapore. 
  	E-mail: feng.y@nus.edu.sg.
  \item
  	Minfeng Zhu is with Zhejiang University. 
  	E-mail: minfeng\_zhu@zju.edu.cn.
  \item*
  	Haoyu Tian and Yingchaojie Feng contributed equally to this work.
  \item$\dag$
  	Minfeng Zhu, Yingchaojie Feng and Wei Chen are the corresponding authors.
}

\abstract{
  The advent of Retrieval-Augmented Generation (RAG) has significantly enhanced the ability of Large Language Models (LLMs) to produce factually accurate and up-to-date responses. However, the performance of a RAG system is not determined by a single component but emerges from a complex interplay of modular choices, such as embedding models and retrieval algorithms. This creates a vast and often opaque configuration space, making it challenging for developers to understand performance trade-offs and identify optimal designs. To address this challenge, we present RAGExplorer, a visual analytics system for the systematic comparison and diagnosis of RAG configurations. RAGExplorer guides users through a seamless macro-to-micro analytical workflow. Initially, it empowers developers to survey the performance landscape across numerous configurations, allowing for a high-level understanding of which design choices are most effective. For a deeper analysis, the system enables users to drill down into individual failure cases, investigate how differences in retrieved information contribute to errors, and interactively test hypotheses by manipulating the provided context to observe the resulting impact on the generated answer. We demonstrate the effectiveness of RAGExplorer through detailed case studies and user studies, validating its ability to empower developers in navigating the complex RAG design space. Our code and user guide are publicly available at \url{https://github.com/Thymezzz/RAGExplorer}.
}


\teaser{
  \centering
  \includegraphics[width=\linewidth, alt={An overview of our visual analytic system.}]
  {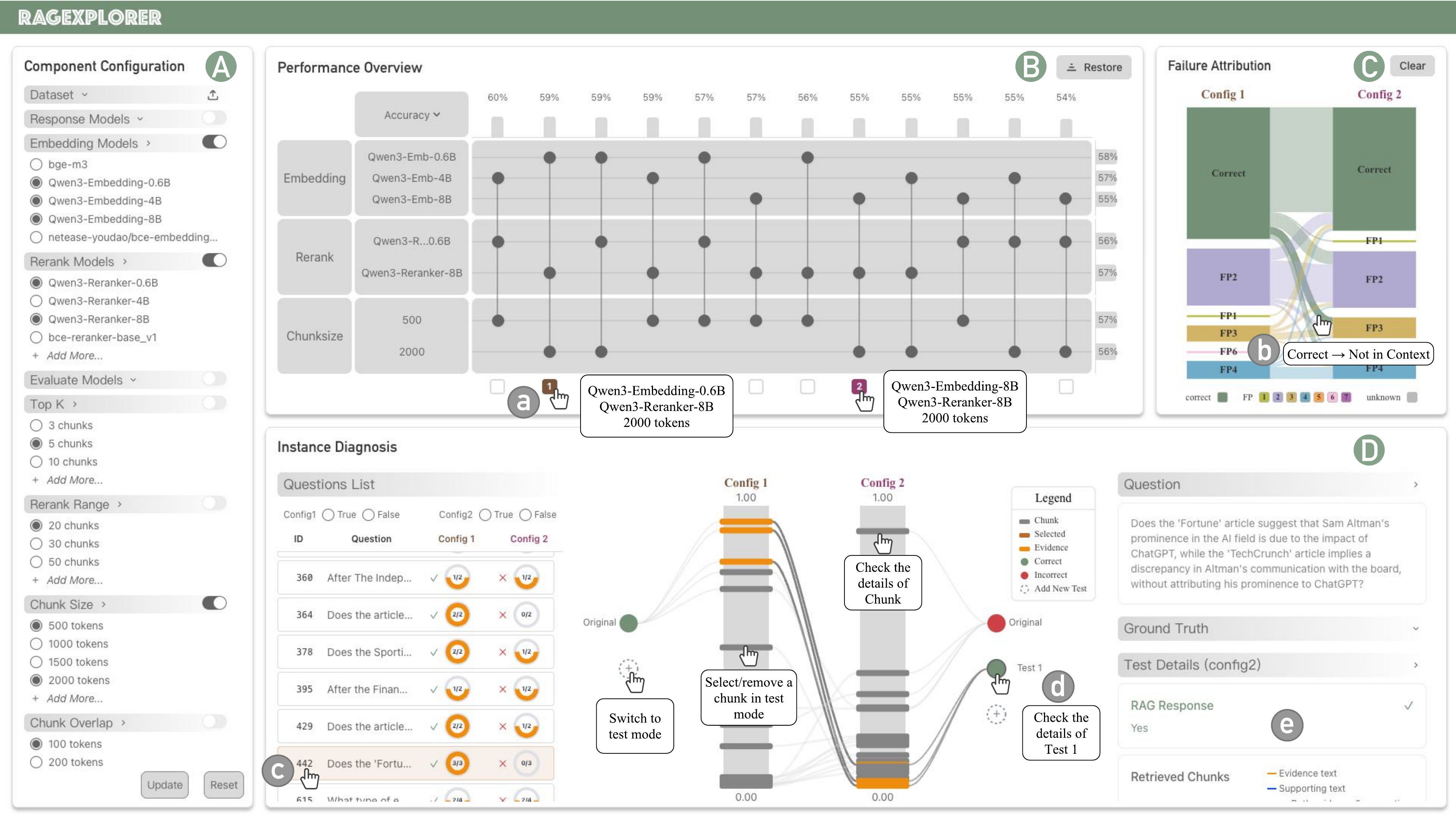}
  \caption{%
    An overview of our visual analytic system. Our system integrates four coordinated views: Component Configuration View (A) allows users to define an experimental space by selecting options for each RAG component. Performance Overview View (B) uses a matrix and coordinated bar charts to display the ranked performance of all configurations, allowing users to identify top performers and assess the average impact of each component choice. The Failure Attribution View (C) visualizes the flow of questions between different failure attribution when comparing two RAG configurations, allowing users to see exactly how a design change impacts RAG's failure points. The Instance Diagnosis View (D) uses a three-panel layout, allowing users to select questions from a list (c), use a central dual-track view to visually compare contexts and interactively test hypotheses (d), and inspect their full text in a details panel (e). %
  }
  \label{fig:teaser}
}

\renewcommand{\manuscriptnotetxt}{}

\graphicspath{{figs/}{figures/}{pictures/}{images/}{./}} 

\usepackage{tabu}                      
\usepackage{booktabs}                  
\usepackage{lipsum}                    
\usepackage{mwe}                       

\usepackage{mathptmx}                  
\usepackage{amsmath}

\usepackage{algorithm}
\usepackage[noend]{algpseudocode}
\usepackage{color}

\usepackage{xcolor} 
\definecolor{revisioncolor}{rgb}{1.0, 0.9, 0.9}

\ifdefined\ShowRevisions
    \newcommand{\wz}[1]{\textcolor{red}{#1}}
    \newcommand{\lhx}[1]{\textcolor{red}{#1}}
    \newcommand{\thy}[1]{\textcolor{red}{#1}}
    \newcommand{\fycj}[1]{\textcolor{red}{#1}}
\else
    \newcommand{\wz}[1]{{#1}}
    \newcommand{\lhx}[1]{{#1}}
    \newcommand{\thy}[1]{#1}
    \newcommand{\fycj}[1]{#1}
\fi

\usepackage[textsize=tiny]{todonotes}

\begin{document}


\firstsection{Introduction}
\maketitle
\vspace{-4pt}
The rapid development of Large Language Models (LLMs) has significantly advanced the capabilities of artificial intelligence. However, their reliance on static training data limits their effectiveness in knowledge-intensive tasks that require timely information, often resulting in factually inaccurate or outdated responses, a phenomenon commonly known as ``hallucination.'' The Retrieval-Augmented Generation (RAG) paradigm~\cite{lewis2020retrieval} was developed to address this limitation. RAG systems dynamically retrieve relevant documents from external knowledge sources to augment the LLM's context, thereby improving the generated content's accuracy, relevance, and reliability.

Despite its promise, the performance of a RAG system is not determined by a single component but emerges from a complex interplay of modular choices, such as document chunking strategies, embedding models, and reranking algorithms. This creates a vast and often opaque configuration space, where the interdependencies between components are difficult to decipher. Consequently, model developers face a critical challenge: \textit{How can they systematically navigate this vast configuration space to understand the performance trade-offs of different design choices and effectively diagnose the root causes of failures?}

Existing approaches offer only partial solutions. While formal evaluation methodologies \cite{simon2024methodology} and automated reporting tools \cite{cohen2025ragxplain} provide valuable metrics and summaries, their static nature hinders intuitive, user-centered exploration of performance data. More importantly, recent visual analytics tools like RAGTrace \cite{cheng2025ragtrace}, RAGViz \cite{wang2024ragviz}, and RAGGY \cite{lauro2025raggy}, though valuable for debugging, are fundamentally designed to diagnose a single RAG pipeline. They focus on the internal workings of one system, rather than enabling the crucial comparative analysis needed to understand performance trade-offs across different configurations. This leaves a significant gap in supporting the holistic optimization of RAG systems.

To bridge this gap, we introduce RAGExplorer, a visual analytics system designed to facilitate the in-depth comparison and diagnosis of RAG configurations, guiding model developers toward more effective designs. The analytical workflow begins in the Component Configuration View (\thy{\cref{fig:teaser}A}), where users define the set of configurations for evaluation. The Performance Overview View (\thy{\cref{fig:teaser}B}) then presents the results in a matrix layout, providing a broad overview that allows users to survey all configurations and \fycj{identify key configurations that can lead to significant performance change.} From there, users can select two configurations for a side-by-side comparison in the Failure Attribution View (\thy{\cref{fig:teaser}C}). This view uses a Sankey diagram to visually articulate how failure patterns shift between them, enabling users to quantify the impact of the design changes and form concrete hypotheses about the root causes of performance differences. 
To validate these hypotheses, users can drill down into the Instance Diagnosis View (\thy{\cref{fig:teaser}D}) for in-depth investigation. It features a novel Dual-Track Context Comparison to visually pinpoint differences in retrieved documents and rankings, alongside an interactive sandbox. In this sandbox, users can directly manipulate the context—for example, by removing a suspected distractor chunk—and immediately observe the impact on the generated answer. This allows them to confirm the root cause of a specific failure.

The main contributions of this paper are:
\begin{enumerate}
    \item A novel comparative diagnosis workflow that shifts the focus of RAG analysis from single-pipeline debugging to the systematic evaluation of multiple configurations.
    \item RAGExplorer, an integrated visual analytics system featuring novel designs for visualizing failure pattern shifts and interactively verifying the root causes of failures.
    \item Two case studies and one expert interview that demonstrate that RAGExplorer effectively helps model developers understand complex performance trade-offs and make informed design decisions.
\end{enumerate}

\section{Related work}
\label{sec:related work}
The related work involves the optimization and evaluation of RAG systems and the visualization for understanding NLP models.

\subsection{Optimization and Evaluation of RAG Systems}
Optimizing and evaluating RAG is a key research area focused on navigating its vast configuration space and diagnostic complexity~\cite{gao2023Retrieval, yu2024Evaluation, gupta2024comprehensive, fan2024survey, ShuyueWEI2025}. To this end, researchers have proposed numerous optimization techniques. One line of work introduces architectural innovations, ranging from flexible modular frameworks~\cite{gao2024modular} to agentic frameworks~\cite{ravuru2024agentic,singh2025agentic} that incorporate mechanisms for self-reflection~\cite{asai2024self} and correction~\cite{yan2024corrective}. Others leverage structured knowledge via knowledge graphs~\cite{li2025cot, edge2024local, peng2024graph}  or design minimalist frameworks for smaller models~\cite{fan2025minirag}. Another line of work focuses on component-level optimizations. These enhance retrieval with techniques like metadata filtering~\cite{poliakov2024multi} and chunk filtering ~\cite{singh2024chunkrag}, or refine generation through methods such as data synthesis for model finetuning~\cite{chen2023few, Yu2024Chain} and credibility-aware attention~\cite{pan2024contexts, Deng2025CrAM}. Furthermore, interactive tools like FlashRAG~\cite{jin2025flashrag} and RAGGY~\cite{lauro2025raggy} accelerate the development cycle. However, these works focus on how to build a better RAG, not on why one configuration outperforms another.

Answering this ``why'' requires effective evaluation and failure analysis. Theoretical work has identified seven common failure points~\cite{barnett2024seven} and highlighted LLM limitations, such as performance degradation with more documents~\cite{levy2025doc} and the ``Lost in the Middle'' effect~\cite{liu2024lost}. Building on this, various evaluation methods have emerged. Some offer reference-free automated evaluation~\cite{es2024ragas}, while others provide fine-grained diagnostics by checking claim-level entailment~\cite{ru2025ragchecker} or assessing document contributions~\cite{salemi2024evaluating, alinejad2024evaluating, wang2024ragviz}. Large-scale benchmarks like RAGBench~\cite{friel2025ragbench} provide comprehensive standards, supplemented by specialized tests for challenges like unanswerability~\cite{peng2025unanswerability} and unified comparisons for long context~\cite{gao2025uniah} and multi-hop queries~\cite{tang2024multihop, yang2018hotpotqa}.

To complement these evaluation approaches, several methodologies and tools have emerged to support RAG system analysis. Simon et al.~\cite{simon2024methodology} propose a systematic comparative methodology for RAG evaluation, though their tabular presentation format limits intuitive insight extraction. Meanwhile, tools like RAGXplain~\cite{cohen2025ragxplain} focus on automated suggestions rather than enabling developer-led cross-configuration comparison. Building on this foundation, RAGExplorer integrates the systematic comparative methodology~\cite{simon2024methodology} with multi-granularity failure attribution analysis, incorporating established failure points~\cite{barnett2024seven} and comparisons against documents selected by user to enable hypothesis-driven debugging.

\subsection{Visualization for Understanding NLP Models}
Visual analytics is an essential tool for understanding the ``black box'' nature of Natural Language Processing (NLP) models and systems~\cite{hohman2019visual, liu2017towards, yang2024foundation, brath2023role, ZHOU2025A1, YE202443, WangWHWHXC23}. These visual analytics approaches can be categorized into two paradigms: single-instance analysis and comparative analysis.

Single-instance analysis focuses on understanding the behavior of an individual model or system workflow. Strobelt et al. visualized the computational mechanisms of traditional NLP models~\cite{strobelt2018lstm, strobelt2019seq}, followed by other researchers using various types of visualizations to interpret attention or neuron distributions of modern NLP models~\cite{vig2019bertviz, hoover2020exbert, deRose2020attentionfa, wang2021dodrio, yeh2023attentionviz, kahng2017cti}. Other single-instance approaches allow users to interactively explore model behavior by performance evaluation~\cite{amershi2015modeltracker} or prompt-based methods~\cite{coscia2023knowledgevis, wang2024commonsensevis, strobelt2022interactive} for LLMs. This analytical paradigm also extends to explaining RAG systems. These works visually interpret specific components within a single RAG workflow, including the knowledge corpus~\cite{yu2024hints}, context utilization~\cite{wang2024ragviz}, the end-to-end pipeline ~\cite{wang2025xgraphrag, cheng2025ragtrace}, and output factuality~\cite{yan2024knownet}.

Comparative analysis evaluates systems by highlighting the differences between them. Early applications of this approach include comparing classifiers by exploring performance on data subsets~\cite{gleicher2020boxer} or using discriminator models~\cite{wang2022learning}. In NLP, comparative techniques have been used to visualize structural differences in embedding spaces~\cite{boggust2022embedding} and to compare textual outputs from various LLMs given the same prompt~\cite{strobelt2021lmdiff, arawjo2024chainforge}. More recently, LLM Comparator~\cite{kahng2025comparator} was developed to support interactive, side-by-side evaluation of LLM outputs. Similar comparative techniques are also applied to study the behavioral impacts of model compression on LLMs~\cite{boggust2025compress}. 
Departing from prior work, RAGExplorer introduces the first visual analytics system for the comparative evaluation of multiple RAG configurations. Our approach adapts the side-by-side comparison methodology to the full system level, revealing the performance impact of different component choices.

\section{Design Process}
In this section, we first illustrate the background on RAG, then distill key challenges and design requirements from a formative study with domain experts, followed by a detailed data description.

\subsection{Background on RAG}
\label{sec:background}
RAG is a paradigm that enhances LLMs by dynamically grounding them in external knowledge sources. This approach is designed to mitigate common LLM failure modes such as factual inaccuracies (``hallucinations'') and reliance on outdated internal knowledge. The performance of a RAG system is highly dependent on the configuration of its modular pipeline, which consists of the following core stages:

\vspace{4pt}
\noindent
    \textbf{Indexing.} A knowledge corpus is processed and structured for efficient search. This involves segmenting documents into text chunks, a process governed by parameters like \texttt{Chunk Size} and \texttt{Chunk Overlap}. These settings control the information's granularity and influence the trade-off between context completeness and retrieval precision.

\vspace{4pt}
\noindent
    \textbf{Encoding.} Text chunks are transformed into high-dimensional numerical vectors using an embedding model. These embeddings capture the semantic meaning of the text, enabling effective similarity comparisons between a query and the document chunks. Models for this task include open-source options like \texttt{bge-m3} and API-based services like \texttt{text-embedding-3-small}.

\vspace{4pt}
\noindent
    \textbf{Retrieval.} Given a user query, the system encodes it and performs a vector similarity search to find the most relevant chunks. The number of initial chunks retrieved is set by the \texttt{Top K} parameter. Optionally, a reranker can be applied to this initial set. Rerankers (e.g., \texttt{BGE-reranker-v2-m3}) use a cross-encoder architecture to more precisely re-evaluate the relevance of each chunk to the query, yielding a better-ordered context.

\vspace{4pt}
\noindent
    \textbf{Generation.} The final retrieved context and the original query are formatted into a prompt and passed to a generator LLM. The LLM's task is to synthesize this information into a coherent, contextually-grounded response. This role can be filled by a range of models, such as \texttt{Gemini 2.5 Flash}, \texttt{GPT-4o-mini}, or \texttt{DeepSeek V3}.

\subsection{Formative Study}
\label{sec:formativeStudy}
To better support analyzing and diagnosing RAG processes, we conduct a formative study with four domain experts. Two of them (E1-2) are model practitioners from a local technology company who focus on improving the robustness and reliability of LLM.
The other two experts (E3-4) are academic researchers who have studied RAG technology for several years and have published research papers at relevant academic conferences.
During our regular meetings, we discuss the traditional pipeline of RAG evaluation and diagnosis, analyze the pain points within the process, and distill the design requirements of our system to address these problems.

Through these collaborative sessions, we synthesized the experts' feedback on the traditional RAG evaluation workflow. We observed that despite their different backgrounds and task scenarios, the experts consistently faced a common set of challenges when relying on standard tools like code notebooks and spreadsheets. 
We summarized three primary challenges as follows.

\begin{enumerate}[label=\textbf{C\arabic*}]
    \item \textbf{Lack of a Holistic Performance Overview.} The development of RAG systems involves tuning numerous parameters, creating a vast and complex configuration space. Typically, users evaluate different parameter combinations and present the performance metrics in spreadsheets. However, this format may hinder users from grasping the global performance landscape and discovering key patterns, leading to the risk of overlooking optimal configurations or failing to understand key parameter interactions.
    
    \item \textbf{Difficulty in Attributing Performance Differences.} While aggregate metrics like accuracy are useful for overall evaluation, they fail to explain why one configuration outperforms another. A higher score can mask the introduction of new, critical failures, obscuring the inherent trade-offs of a design choice. Lacking support for systematic diagnosis, users must resort to laborious manual review to investigate the root causes, which severely hinders efficient optimization.
    
    \item \textbf{Inefficient Instance-Level Debugging.} Pinpointing the root cause of individual failures often requires a deep dive into the retrieved context to investigate issues. In current practice, users typically need to manually inspect raw text within a plain text environment and repeatedly modify the original text to test the hypothesis. This fractured loop between analysis and experimentation can be time-consuming and may impose a significant cognitive load, potentially disrupting the analytical flow.
\end{enumerate}

\subsection{Design Requirements}
\label{Sec:requirents}
Based on the challenges identified in our formative study, we derive and iteratively refine a set of design requirements that guide the development of our visual analysis system. These requirements aim to address the limitations of traditional workflows by enabling a more systematic and efficient analytical process.

\begin{enumerate}[label=\textbf{R\arabic*}]
    \item \textbf{Provide a Holistic Performance Overview.} The system should help developers grasp the global performance landscape across all configurations \fycj{(\textbf{C1})}, enabling users to quickly identify high-performing or anomalous configurations and discover high-level performance patterns. Moreover, it should support the assessment of a single design choice (e.g., a specific model or parameter). This helps users understand the contribution of individual components to the system's overall performance.

    \item \textbf{Facilitate a Diagnostic Comparison of Configurations.} Aggregate performance metrics often obscure the nuanced trade-offs inherent in design choices. To help developers understand why one configuration outperforms another \fycj{(\textbf{C2})}, the system must support in-depth diagnostic comparisons. This allows developers to precisely attribute performance gains or losses to specific failure modes, revealing the inherent trade-offs of a design decision. For instance, it can show how a reduction in one type of failure cases may come at the expense of introducing another. By doing so, the system enables a balanced assessment of each configuration's relative strengths and weaknesses.

    \item \textbf{Streamline Instance-Level Debugging.} To mitigate the fractured process of diagnosing individual failures \fycj{(\textbf{C3})}, the system should provide an integrated sandbox for interactive debugging. It should allow users to directly and intuitively perturb the context of failure cases to conduct rapid what-if analysis. For instance, users can interactively choose different chunks as context to identify the noise information in different chunks that lead to failure. This enables users to efficiently verify the failure causes.
\end{enumerate}

\subsection{Data Description}
\label{sec:dataDescription}
Our evaluation is conducted on the \texttt{MultiHop-RAG dataset}~\cite{tang2024multihop}, which comprises 2,556 questions derived from an English news corpus. A key characteristic of this dataset is its focus on multi-hop questions. Each question necessitates retrieving and reasoning over evidence scattered across multiple documents to formulate a correct answer, \fycj{which is a challenging task for general RAG pipelines}. The dataset's inherent complexity, including both multi-hop reasoning and instances with insufficient evidence, provides a testbed for diagnosing the nuanced failure modes of RAG systems. 
The dataset also includes ground-truth \texttt{evidence} annotations for the exact evidence sentences required to answer each question. Inspired by LLM self-explanation techniques, to enable granular, automated analysis of the generation process, we mandate a structured output format. The LLM is prompted to return its response as a JSON object containing two mandatory keys: \texttt{supporting\_sentences}, an array of the exact evidence sentences used for reasoning, and \texttt{final\_answer}, a concise answer constrained to a maximum of three words.

\section{Framework}
\begin{figure*}[t]
	\centering
	\includegraphics[width=\textwidth]{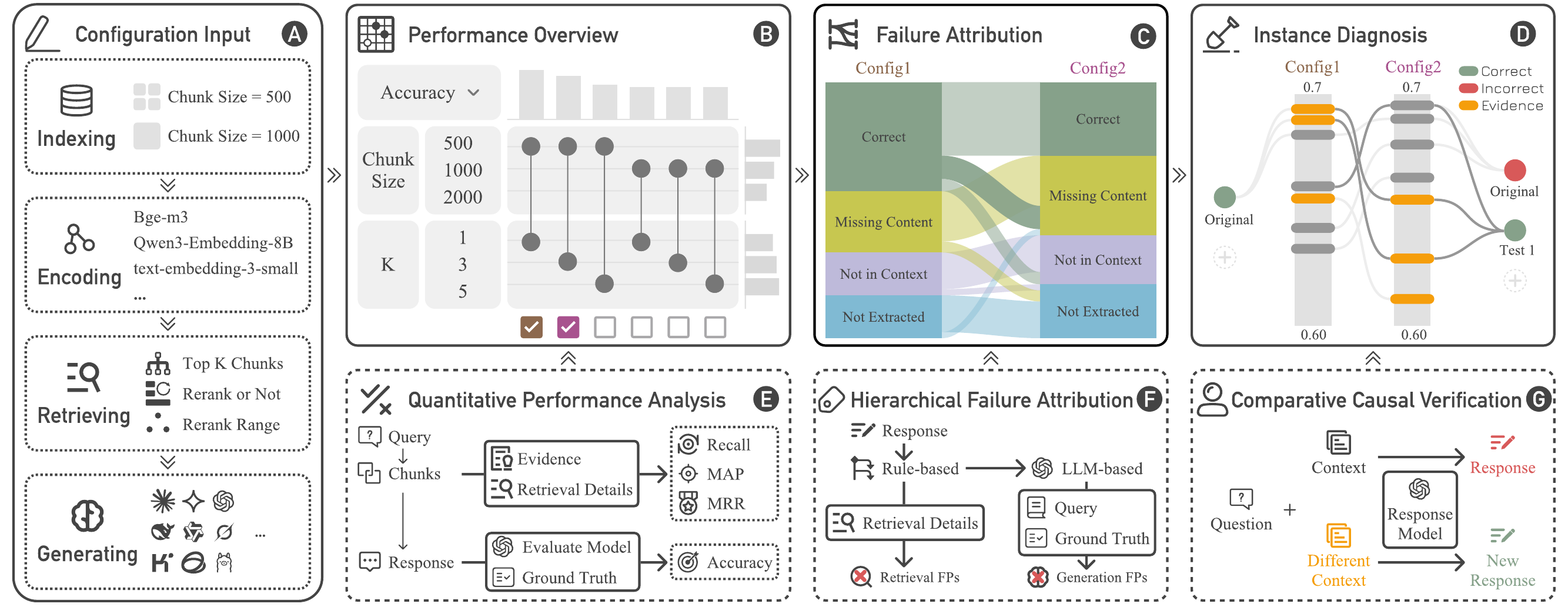}
	\caption{RAGExplorer helps users optimize RAG pipelines through four main stages. After a user (A) defines a configuration space, the system generates (B) a performance overview based on (E) retrieval metrics (\texttt{Recall}, \texttt{MRR}, \texttt{MAP}) or \texttt{Accuracy}. Based on this overview, users can (C) select two configurations to compare, for which (F) an automated algorithm provides a hierarchical attribution of their failure points. This allows users to (D) analyze the context similarity distribution. They can also (G) modify the context to regenerate answers and verify the impact of a specific instance.}
    \vspace{-4pt}
	\label{fig:rag_pipeline}
\end{figure*}

The core of our framework is a three-stage analytical methodology designed to move from high-level metrics to specific root causes (\cref{fig:rag_pipeline}). The process begins with (B) Quantitative Performance Benchmarking that computes a suite of metrics for system evaluation. To diagnose the underlying reasons for performance gaps, we employ (C) a Hierarchical Failure Attribution algorithm that diagnoses and categorizes the root causes of failures. Finally, the framework facilitates (D) Interactive Causal Verification, where users can test hypotheses by directly intervening in the reasoning process.

\subsection{Quantitative Performance Analysis}
To quantitatively assess RAG performance, our methodology(\cref{fig:rag_pipeline}E) evaluates the distinct contributions of the retrieval and generation stages. We adopt a set of standard metrics from information retrieval, inspired by recent surveys on RAG evaluation~\cite{yu2024Evaluation}.

For the retrieval stage, we measure the efficacy of evidence retrieval using three established metrics:

\begin{itemize}
\item \textbf{Recall@k} quantifies the proportion of relevant documents captured within the top-k results, indicating if the necessary evidence was available to the generator. It is defined as $\text{Recall}@k = \frac{|\text{Relevant} \cap \text{Retrieved}_k|}{|\text{Total Relevant}|}$.

\item \textbf{Mean Reciprocal Rank (MRR)} assesses the system's ability to rank the \textbf{first} relevant document as high as possible. This is critical for efficiency and user trust, calculated as $\text{MRR} = \frac{1}{|Q|} \sum_{i=1}^{|Q|} \frac{1}{\text{rank}_i}$, where $\text{rank}_i$ is the rank of the first relevant document for the \textit{i}-th query.

\item \textbf{Mean Average Precision (MAP)} provides a holistic measure of ranking quality. Unlike \texttt{MRR}, \texttt{MAP} considers the rank of \textbf{all} relevant documents, rewarding systems that place more correct documents at the top of the list. It averages the precision at each relevant document's position across all queries.

\end{itemize}

For the end-to-end generation stage, we measure the quality of the final output using the following metric:
\begin{itemize}
\item \textbf{Factual Correctness}: This metric quantifies the factual accuracy of the generated \texttt{final\_answer}. We adopt the LLM-as-a-Judge paradigm, a methodology validated by automated evaluation frameworks like RAGAS~\cite{es2024ragas}. An adjudicator model is employed to assess whether the generated answer is semantically equivalent to the ground truth. The final score is the proportion of answers evaluated as correct.
\end{itemize}

\begin{figure}[t]
    \centering
    \includegraphics[width=\linewidth]{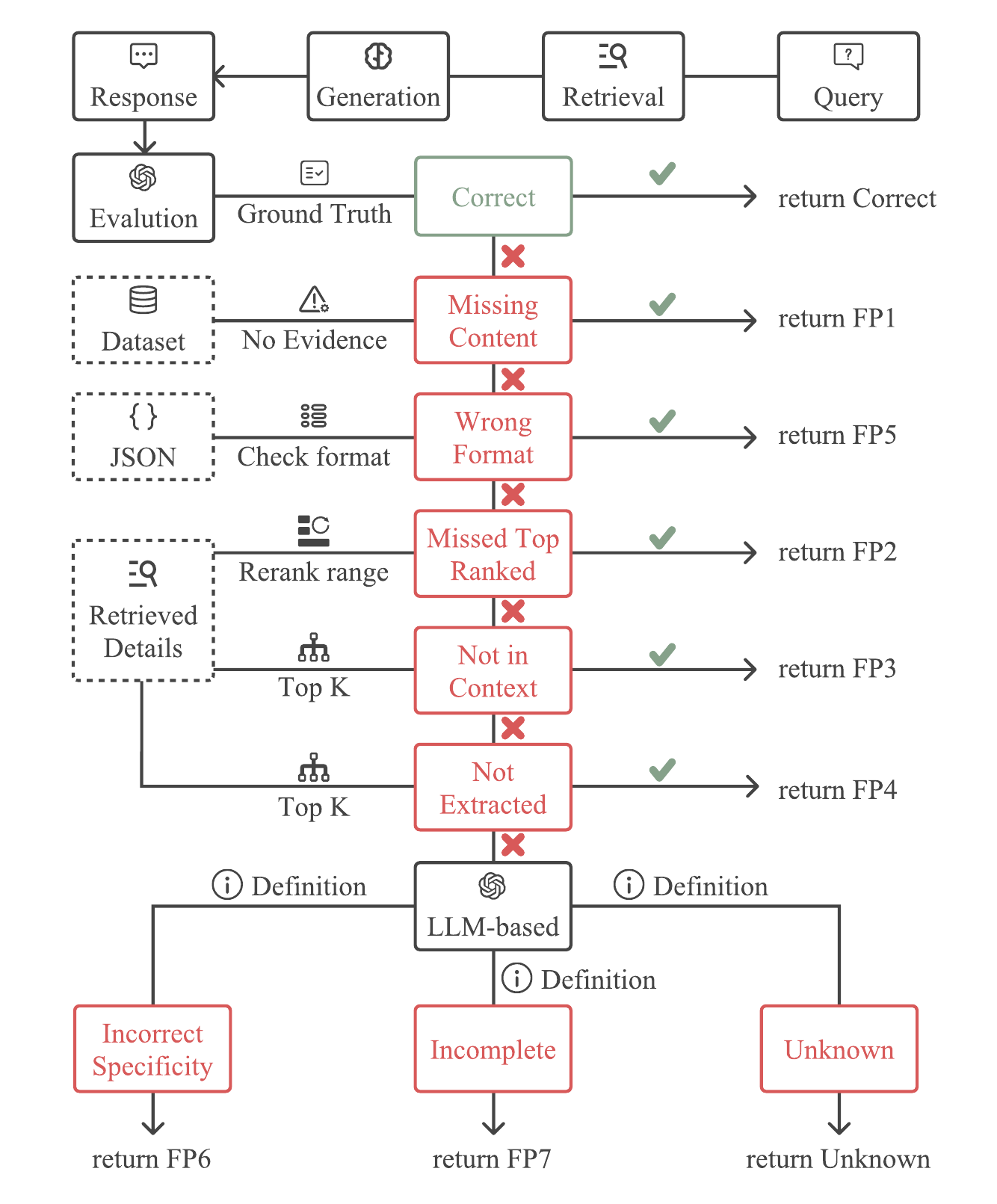}
    \caption{A hierarchical algorithm for diagnosing RAG failure points. It applies a prioritized cascade of checks to assign a single, primary failure point to each incorrect answer.}
    \vspace{-4pt}
    \label{fig:hierachical}
\end{figure}

\subsection{Hierarchical Failure Attribution}
\label{sec:hierachical}

To systematically diagnose RAG failures, this work introduces an automated, hierarchical failure attribution algorithm (\cref{fig:hierachical}).

\lhx{
Our method is inspired by the catalogue of failure points with Barnett et al.~\cite{barnett2024seven}. It begins by identifying correct answers. It labels them \texttt{correct} and stops. For all incorrect answers, the algorithm proceeds through a sequence of checks to assign a single, primary failure point (\texttt{FP}).
}

\lhx{
The first check identifies \texttt{Missing Content (FP1)}, which assesses the system's handling of unanswerable queries. For questions where the ground truth is ``insufficient information,'' the RAG system is prompted to produce the same output. If the system generates any other answer, this mismatch is categorized as \texttt{FP1}.
The next check is for \texttt{Wrong Format (FP5)}. It evaluates the RAG's instruction-following capability. Our parser can often handle malformed JSON and still extract a \texttt{final\_answer}. But the system logs this case as \texttt{FP5}. 
}

\lhx{
If the initial checks pass, the analysis proceeds to the retrieval stage. The algorithm examines the evidence within the ``rerank range.'' A failure is categorized as \texttt{Missed the Top Ranked Documents (FP2)} if the proportion of ground-truth evidence sentences in this range falls below a 70\% threshold. While a 100\% threshold is theoretically ideal, we adopted 70\% based on empirical observations that RAG systems can often output correct answers with partial evidence. This may be due to the synthetic nature of the dataset or the existence of valid reasoning paths not captured in the ground truth.} 
\thy{The system might find sufficient evidence (>=70\%) in the rerank range. But the final top-k context passed to the LLM may be insufficient. We apply the same 70\% threshold. This failure is classified as \texttt{Not in Context (FP3)}.} 

\lhx{
Failures not caught in the preceding stages are attributed to the generation module. Among these, \texttt{Not Extracted (FP4)} is identified under a strict condition: the final context provided to the LLM contained 100\% of the required ground-truth evidence, yet the answer was still incorrect. This indicates the generator failed to either identify the evidence or correctly reason with the complete information provided.
}

\lhx{For more nuanced generation failures, we employ an LLM-based adjudicator. The adjudicator classifies these cases into \texttt{Incorrect Specificity (FP6)} (e.g., answering ``France'' when the question asks for ``Paris'') or \texttt{Incomplete (FP7)} (missing parts of the answer). To perform this classification, the LLM judge is provided with a prompt containing the failure definitions, the original \texttt{query}, the \texttt{ground\_truth} answer, the model's \texttt{final\_answer}, and its \texttt{raw\_JSON\_response}. An \texttt{Unknown} category is included for cases the adjudicator finds ambiguous or difficult to classify.}

\subsection{Comparative Causal Verification} 
To systematically investigate how the composition of a retrieved context influences the final generated answer, we employ a method of comparative causal verification (\cref{fig:rag_pipeline}G). This process begins by establishing a content-based correspondence between the chunk sets of two configurations, $C_A$ and $C_B$, which is essential when their underlying chunking strategies \thy{or embedding models} may differ. A pairwise similarity score is computed for every chunk $c_i \in C_A$ and $c_j \in C_B$ using the Jaccard index to quantify their textual overlap:
\begin{equation}
J(c_i, c_j) = \frac{|W_i \cap W_j|}{|W_i \cup W_j|},
\end{equation}
where $W_i$ and $W_j$ represent the sets of unique words in chunks $c_i$ and $c_j$. The resulting similarity matrix reveals the informational overlap and divergence between the two contexts. 
\thy{Based on the matrix, we retain only chunk pairs whose similarity exceeds a user-adjustable threshold(default 0.3), balancing recall of nuanced overlaps against the precision of stronger matches. Having these connections, we can conduct a causal contrastive analysis along the paths of context preservation, substitution, or omission, tracing which contextual changes truly drive the differences in generated answers.}

Building on this comparative analysis, causal hypotheses are tested via context perturbation. Let $f_{gen}$ be the generation function and $C_{orig}$ be an original context that yields an answer $Ans_{orig} = f_{gen}(C_{orig})$. A new, curated context, $C_{pert}$, is then constructed by modifying the composition of $C_{orig}$. Such modifications can include removing chunks hypothesized to be irrelevant distractors or substituting chunks with high-similarity counterparts from the alternate configuration. The verification is completed by generating a new answer, $Ans_{pert} = f_{gen}(C_{pert})$, and observing the change relative to $Ans_{orig}$. For instance, if an incorrect answer becomes correct after a specific chunk is removed, it provides evidence for a causal link between that chunk and the initial failure. This method enables an empirical validation of hypotheses regarding phenomena like the model's susceptibility to distraction or the ``Lost-in-the-Middle'' \thy{\cite{liu2024lost}} effect.

\section{System Design}
RAGExplorer is a visual analysis system designed to facilitate the systematic comparison and diagnosis of RAG systems. 
We introduce its workflow, followed by detailed descriptions of four visualization views.
\Cref{fig:teaser} shows an overview of RAGExplorer's interface.


\subsection{Workflow}
RAGExplorer follows a multi-stage workflow, moving from a broad overview of system performance to a detailed investigation of individual instances.
The workflow begins from the Component Configuration View (\cref{fig:teaser}A), where the user defines a configuration space for analysis. 
The user specifies candidate models for modular components of RAG systems, including \texttt{response models} (generator), \texttt{embedding models}, and \texttt{rerank models} (reranker), as well as key parameters, such as \texttt{chunk size} and \texttt{overlap}.

Upon evaluating the specified configurations, the user proceeds to the Performance Overview View (\cref{fig:teaser}B) to survey the performance landscape. 
The matrix-based layout enables a systematic overview across all configurations \fycj{(\textbf{R1})}, facilitating the identification of high-performing designs, unexpected outliers, and overarching patterns.
To diagnose the discrepancy between distinct configurations, the user selects the two configurations of interest for a side-by-side comparison in the Failure Attribution View (\cref{fig:teaser}C). 
This view employs a Sankey diagram to visually articulate the precise shifts in instance-level outcomes. 
By tracing the flows between states, the user can quantify the impact of a design change and form concrete hypotheses about the root causes of performance differences \fycj{(\textbf{R2})}.
For example, they might attribute the performance drop to a significant flow of instances from \verb|Correct| to \verb|FP4: Not Extracted| failures, suggesting that while the correct information was retrieved, it was ultimately ignored by the generator.

To validate the hypothesis, the user needs to select the corresponding flow in the Sankey diagram, which filters the Instance Diagnosis View (\cref{fig:teaser}D) to the relevant set of questions. This final view enables an interactive causal verification and debugging loop at the instance level \fycj{(\textbf{R3})}. 
Suspecting that an irrelevant ``distractor'' chunk retrieved alongside the correct evidence is causing the failure, the user can temporarily remove it from the context and regenerate the answer. 
An immediate change in the output from incorrect to correct provides direct, empirical evidence confirming the chunk's causal role in the failure, thus completing the analytical process from high-level observation to validated, instance-level insight.

\subsection{Component Configuration View}
The Component Configuration View (\cref{fig:teaser}A) serves as the entry point of analysis, designed to define a manageable and targeted subset of this space for systematic exploration.
The interface consists of a series of selection panels, one for each modular component of the RAG pipeline (e.g., \texttt{embedding models}) and its key parameters (e.g., \texttt{chunk size}).
Within each panel, the user can select one or more options. 
The system then computes the Cartesian product of these selections to generate the full set of configurations to be evaluated. 
Once the user finalizes their selection and initiates the evaluation, the system executes the pipelines for all questions in the dataset~\cite{tang2024multihop} in parallel to expedite the process.

\subsection{Performance Overview View}
To facilitate a holistic performance assessment (\textbf{R1}), the Performance Overview View (\cref{fig:teaser}B) is designed to support two key analytical tasks: identifying the highest-performing overall configurations and discerning the general effectiveness of individual components. 
We employ a coordinated matrix and multi-chart layout to support analysis.
\thy{
Inspired by UpSet \cite{upset2014}, our design visualizes the intersections among multiple configuration sets, enabling users to intuitively observe how different parameter combinations influence overall performance.
}

The central element of this view is a configuration matrix, where each column represents a unique RAG pipeline and each row corresponds to a specific component choice. 
A glyph at each intersection visually links a configuration to its constituent parts.
This matrix is coordinated with two summary bar chart visualizations: 
\vspace{-1mm}
\begin{itemize}[leftmargin=*]
    \item \textbf{Global performance} visualization is positioned above the matrix. Each bar corresponds to a configuration column, and its length encodes a selected performance metric, including \texttt{Accuracy, Recall, MRR, and MAP}. This provides a ranked overview for quickly identifying the best- and worst-performing pipelines. 
    \vspace{-1.5mm}
    \item \textbf{Component-wise summary bar chart} visualization is positioned to the right of the matrix. Each bar represents a single component choice (e.g., the \texttt{bge-m3 embedding model}), and its length encodes the average performance of all configurations that include that component. This facilitates an assessment of a component's general effectiveness, independent of any single pipeline.
\end{itemize}
\vspace{-1mm}
The user can change the primary metric, which re-sorts the configurations in the global chart and the matrix. 
The principal action is the selection of two configuration columns, which initiates a comparative analysis by populating the Failure Attribution View (\cref{fig:teaser}C).

\subsection{Failure Attribution View}

To facilitate the systematic diagnosis of performance shifts between two configurations (\textbf{R2}), the Failure Attribution View (\cref{fig:teaser}C) moves beyond aggregate metrics to visualize how instance-level outcomes change. 
We employ a Sankey diagram for this purpose, as its flow metaphor is well-suited to representing the transition of questions between discrete states.
The diagram's two columns represent the selected configurations, while the nodes correspond to outcome states: \verb|Correct| or a specific failure category from our hierarchical failure attribution algorithm (Sec ~\ref{sec:hierachical}), such as \verb|FP2: Missed Top Ranked|.
The width of the flows connecting nodes between the two columns is proportional to the number of questions that transitioned from the source state to the target state. 
This encoding provides an immediate visual summary of the impact of a design change, such as a large flow from \verb|FP2| in the baseline to \verb|Correct| in the variant, indicating improved retrieval.

To enable a seamless drill-down from pattern to detail, the view supports two interactions: (1) hovering over a flow or node reveals a tooltip with the exact instance count; (2) clicking on a flow filters the Instance Diagnosis View (\cref{fig:teaser}D) to show only the questions that constitute that specific transition, preserving the analytical context.

\subsection{Instance Diagnosis View}
To support efficient instance-level debugging and hypothesis testing (\textbf{R3}), the Instance Diagnosis View (\cref{fig:teaser}D) integrates three coordinated views for detailed analysis of individual questions.

\vspace{4pt}
\noindent
\textbf{Question List.} 
This view (\cref{fig:teaser}c) is populated based on the user selection in the Failure Attribution View, which filters questions from the dataset.
This list serves as a navigator, where each entry includes a circular glyph that visually encodes the proportion of ground-truth evidence retrieved, offering a pre-attentive cue to retrieval quality.

\vspace{4pt}
\noindent
\textbf{Dual-Track Context Comparison.} 
This view (\cref{fig:teaser}d) enables a direct side-by-side inspection of the retrieved contexts from two configurations.
It presents two parallel vertical tracks, one for each configuration.
Within each track, retrieved text chunks are represented as horizontal bars, with their vertical position determined by their relevance score (top for highly relevant).
Connecting lines between the tracks highlight textually similar chunks, making critical differences—such as presence, absence, or changes in rank—immediately apparent.
\wz{This visualization supports a human-in-the-loop process of hypothesis testing, such as hypothesizing a chunk acting as a distractor, by manually adding or removing the chunk in context (selecting or deselecting the chunk bar in the tracks) and regenerating the answer to observe its causal impact.}

\begin{figure}[t]
    \centering
    \includegraphics[width=\linewidth]{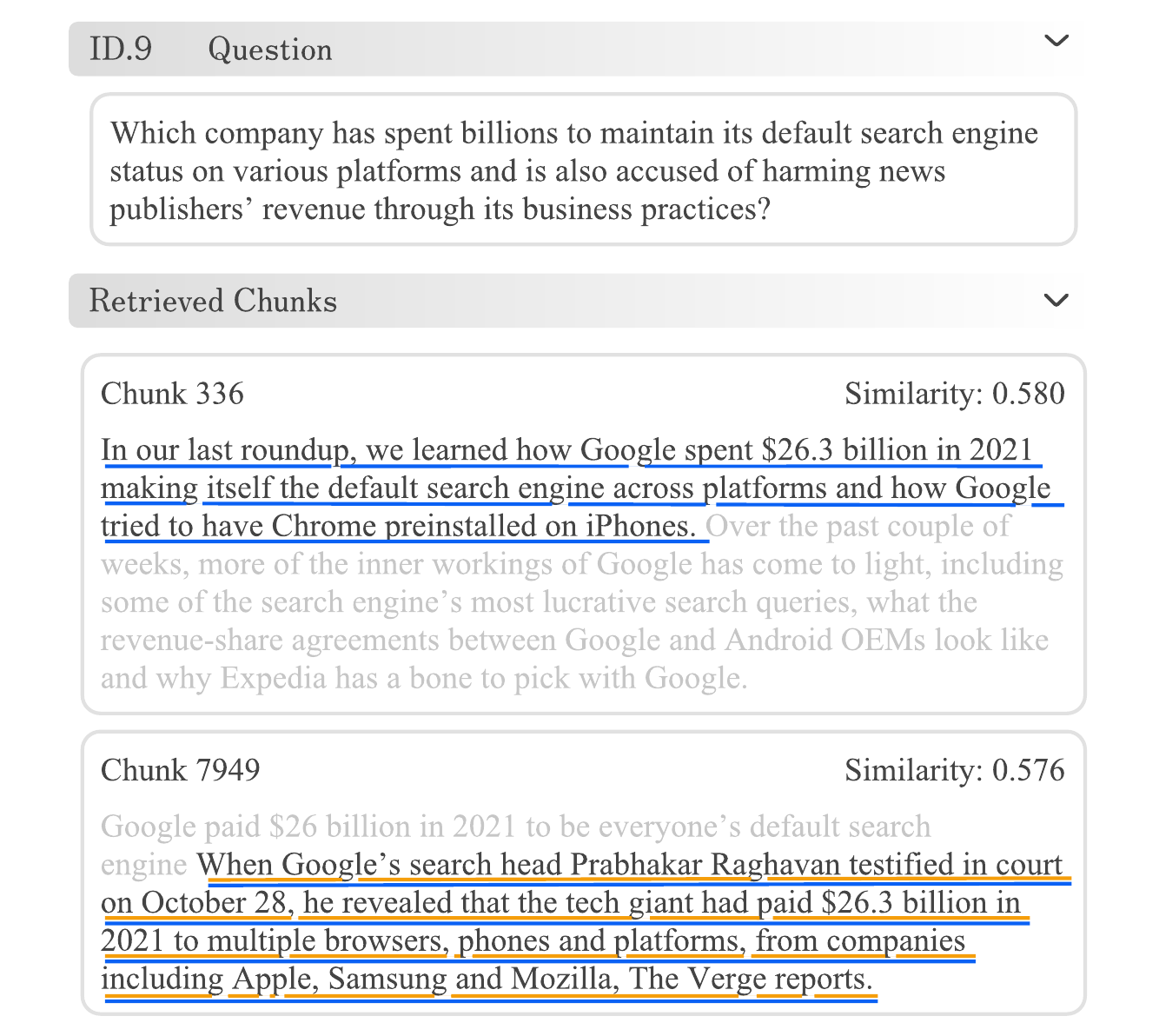}
    \caption{This figure illustrates question ID.9 with retrieved chunk 336 and chunk 7949. Irrelevant text is grayed out, \texttt{supporting\_sentences} are underlined in blue, and \texttt{evidence} is highlighted in orange. These stylings may be combined where the text passages overlap.}
    \vspace{-4pt}
    \label{fig:highlight}
\end{figure}

\vspace{4pt}
\noindent
\textbf{Text Details Panel.}
This view (\cref{fig:teaser}e) on the right supports this inspection by providing the full text of selected item. 
To facilitate evidence tracing, the panel uses targeted highlighting: ground-truth \texttt{evidence} found within a chunk is underlined in orange, while sentences cited in the model's response are underlined in blue (\cref{fig:highlight}).

\section{Case Study}
\label{sec:case study}
\thy{Our case studies were conducted by two experts (E1, E2) from the formative study (\cref{sec:formativeStudy}). Our first case study demonstrates how an expert (E1) uses our system to resolve a contradiction between industry consensus and recent research. Our second case study demonstrates how an expert (E2) uses our system to diagnose a counter-intuitive component failure, revealing how a ``stronger-is-better'' hypothesis can be a performance trap.}

\subsection{Resolving the Overlap Contradiction}

\thy{
\begin{figure*}[htbp]
  \centering
  \includegraphics[width=\textwidth]{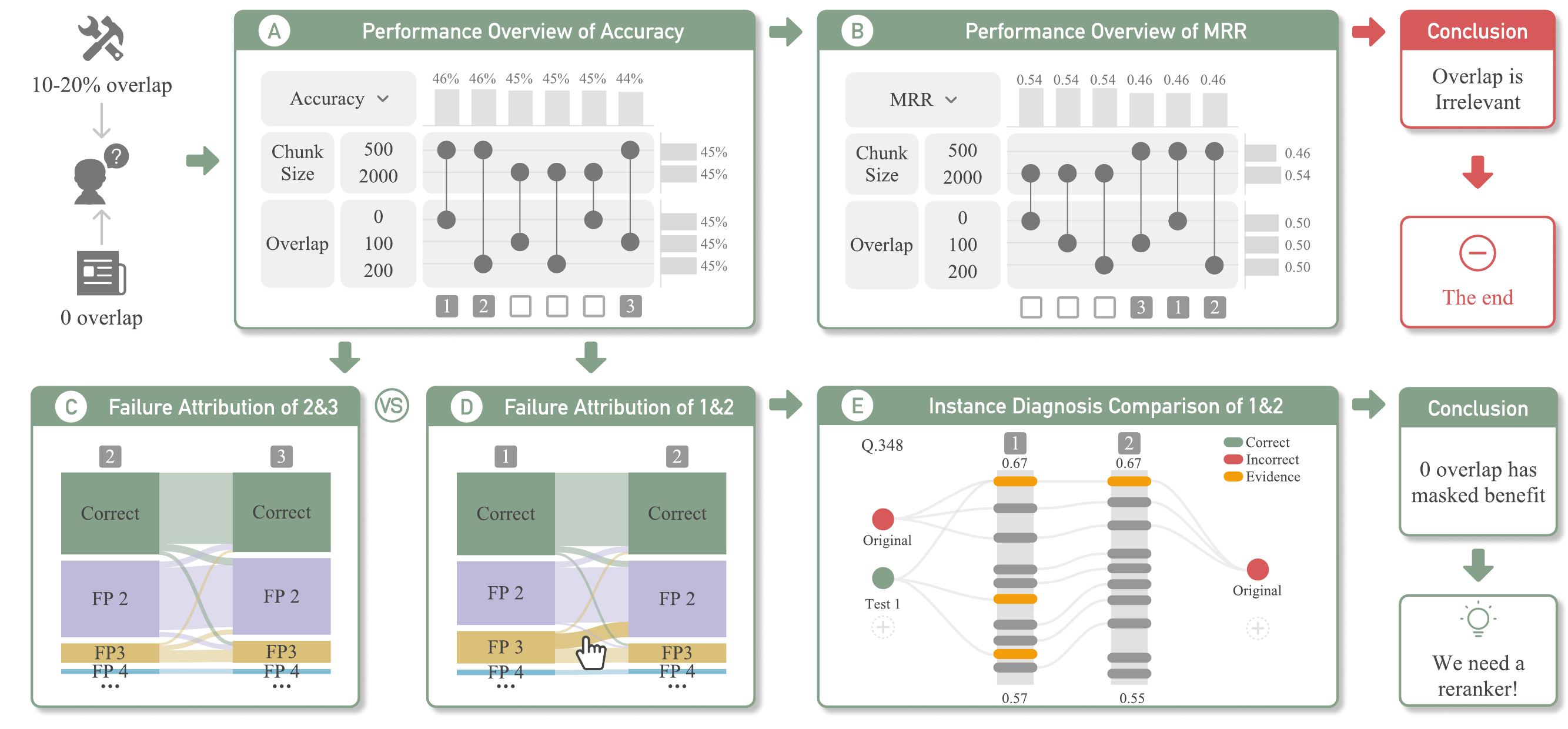}
  \caption{
This case study illustrates how an expert resolves a conflict about \texttt{overlap} using our system.
(A–B) In the Performance Overview of \texttt{Accuracy} and \texttt{MRR}, all \texttt{overlap} settings (0, 100, 200) appear identical, suggesting that \texttt{overlap} is irrelevant.
(C–D) Guided by the Failure Attribution View, the expert finds that \texttt{overlap}=100 and 200 behave similarly, but \texttt{overlap}=0 shows a distinct failure pattern—fewer \texttt{FP2 (missed top-ranked documents)} yet more \texttt{FP3 (not in context)}.
(E) The Instance Diagnosis Comparison of case Q.348 reveals why: \texttt{overlap}=0 retrieves more evidence into the rerank range, though not ranked in top-k. This uncovers a benefit and suggests the next step: adding a reranker.
}
\vspace{-4pt}
\label{fig:case1}
\end{figure*}
}

\thy{
E1 was motivated by a conflict: ``industry consensus'' often suggests a small \texttt{overlap} (e.g., 10–20\% of \texttt{chunk size}) \cite{bestchunkingrag} to avoid semantically splitting text or too much noise. However, some recent research \cite{DBLP:journals/corr/abs-2506-17277} suggests an \texttt{overlap}=0 strategy may be superior. To isolate the parameter's effect, he fixed other variables (\texttt{Top K}, \texttt{embedding model}) and disabled the reranker. He then tested \texttt{overlap}=0 (manually added in View A) against consensus values (100 and 200) across different \texttt{chunk size} (500 and 2000).
}

\thy{ E1's initial results in the Performance Overview (View B) appeared puzzling (\cref{fig:case1}A). They failed to resolve the conflict: the average \texttt{accuracy} scores for all overlap configurations (0, 100, and 200) were nearly identical. The expert decided to check other metrics (\cref{fig:case1}B). He examined retrieval-specific statistics, including \texttt{Recall}, \texttt{MAP}, and \texttt{MRR}, and found these ``simple statistics'' also showed almost no difference between the configurations. At this point, a user relying only on statistical tables would be completely misled. They would wrongly conclude that \texttt{overlap} is an unimportant parameter and that both the consensus and the new research were incorrect, at least for this dataset. }

\thy{ The expert then used the Failure Attribution View (View C) to find the truth hidden beneath the averages. He compared two of the top-ranked configurations (\cref{fig:case1}D): \texttt{overlap}=0, \texttt{chunk size}=500 and \texttt{overlap}=200, \texttt{chunk size}=500. The view immediately revealed a non-obvious finding that all statistical metrics had been completely masked. Although their final scores were nearly identical, their failure patterns were significantly different. The \texttt{overlap}=0 configuration had noticeably fewer \texttt{FP2 (missed top-ranked documents)}. At the same time, it had noticeably more \texttt{FP3 (Not in context)}. To confirm this was a unique property of \texttt{overlap}=0, he also compared \texttt{overlap=100}, \texttt{chunk size}=500, and \texttt{overlap}=200, \texttt{chunk size}=500, and found they were nearly identical in the Failure Attribution View (\cref{fig:case1}C). }

\thy{ To confirm why this failure mode had shifted, the expert drilled down into the Instance Diagnosis View (View D). He selected the main source of the difference: the flow of questions that were \texttt{FP2} in the \texttt{overlap}=200 config but became \texttt{FP3} in the \texttt{overlap}=0 config. The Instance Diagnosis View (\cref{fig:case1}E) provided a visual proof. For these questions, the \texttt{overlap}=0 pipeline often successfully retrieved more groundtruth evidence into the rerank range (which led to fewer \texttt{FP2}). However, he also observed that this new evidence was often ranked low, failing to make the final top-k (thus causing more \texttt{FP3}). }

\thy{ This complete diagnostic process provided the expert with a clear path for an informed improvement. He now understood the contradiction: the overlap=0 pipeline had superior recall (fewer \texttt{FP2}) but its gains were being hidden by a new bottleneck at the next sequential step of RAG (more \texttt{FP3}). His next concrete step was to create a new configuration (\texttt{overlap}=0 combined with a newly added reranker) to exploit this proven recall benefit and fix the identified ranking bottleneck. }

\begin{figure}[htbp]
    \centering
    \includegraphics[width=\linewidth]{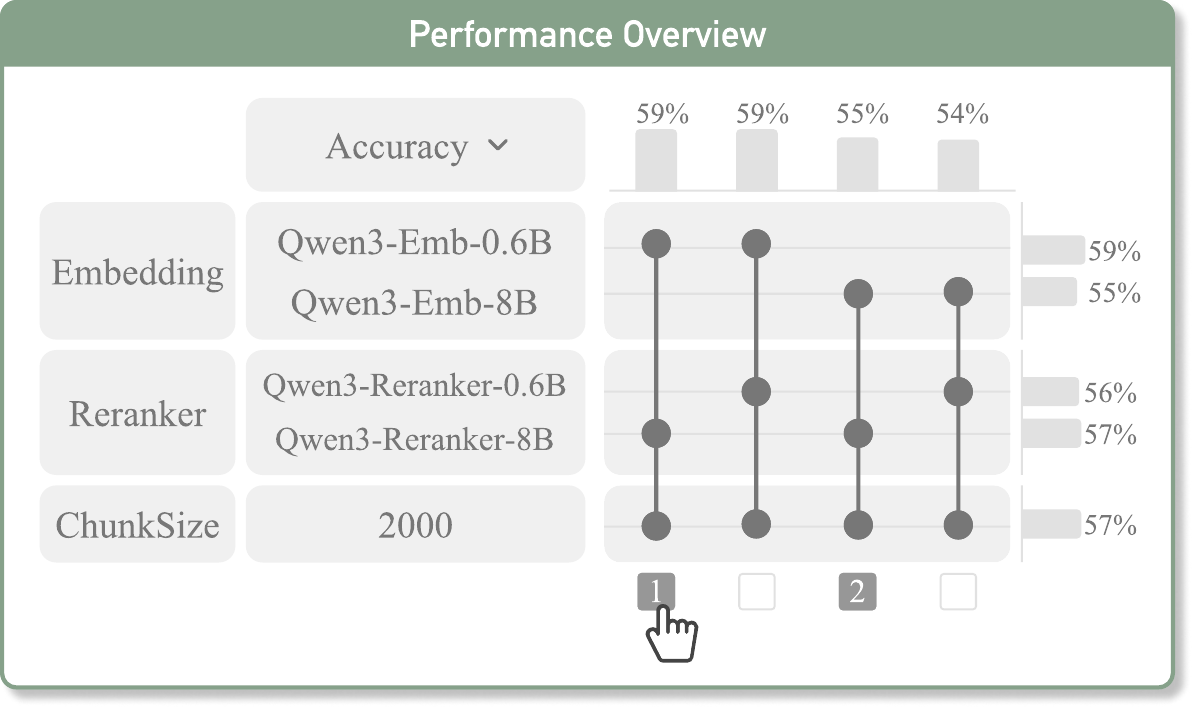}
    \caption{The Performance Overview results for the expert's initial ``stronger-is-better'' hypothesis. The view shows the ``Strongest'' configuration (Config A: \texttt{emb-8B} + \texttt{reranker-8B} + \texttt{chunk size=2000}) achieves 55\% accuracy, while the ``Budget'' configuration (Config B: \texttt{emb-0.6B} + \texttt{reranker-0.6B} + \texttt{chunk size=2000}) achieves 59\% accuracy.}
    \vspace{-4pt}
    \label{fig:case2}
\end{figure}

\thy{
\subsection{Diagnosing Counter-Intuitive Component Failure}
An expert (E2) begins by testing a ``stronger-is-better'' hypothesis. In the Component Configuration View, he selects a range of parameters he wants to compare, including \texttt{Qwen3-Embedding-0.6B/8B}, \texttt{Qwen3-Reranker-0.6B/8B}, and \texttt{chunk size} set to 2000. The system then runs all possible combinations. He expects Config A (\texttt{emb-8B} + \texttt{reranker-8B} + \texttt{chunk size=2000}) to perform best. The Performance Overview (\cref{fig:case2}) reveals a counter-intuitive finding: Config B (\texttt{emb-0.6B} + \texttt{reranker-0.6B} + \texttt{chunk size=2000}) achieves 59\% accuracy, while Config A is 4\% worse, scoring only 55\%. }

\thy{
Why ``stronger'' components lead to a decrease? The expert uses the Failure Attribution View to compare Config A (55\%) with Config B (59\%). The view (\cref{fig:teaser}C) immediately reveals that Config A (\texttt{emb-8B} + \texttt{reranker-8B}) has a massive new bottleneck, with significantly more \texttt{FP3 (Not in context)} and \texttt{FP4 (Not extracted)}. Many of these new failures were questions that Config B had answered correctly. The Instance Diagnosis View confirms why: the \texttt{emb-8B} model's ``over-retrieval'' of similar but distracted chunks is confusing the \texttt{reranker-8B}. For example, in Q.442 (\cref{fig:teaser}D), the Dual-Track Context Comparison shows that while Config B correctly ranked the evidence in the top-k, Config A puts them at the bottom instead.
}

\thy{
This diagnosis leads to the final step of the expert's exploration. The initial ``stronger-is-better'' hypothesis was proven wrong. If \texttt{chunk size=2000} with the most ``powerful'' embedding model brings too much noise and causes more failure points, how about a smaller \texttt{chunk size} with a less powerful embedding model? He thought it could be a promising improvement.
}

\thy{
To test this, he adds the \texttt{emb-4B} model and \texttt{chunk-500} to his Component Configuration View, broadening his search to include mid-range and less-noisy options. After running the new combinations, he returns to the Performance Overview View (\cref{fig:teaser}B). He immediately discovers that his new idea worked. The top-performing configuration is now Config C (\texttt{emb-4B} + \texttt{reranker-0.6B} + \texttt{chunk-500}), which achieves 60\% accuracy.
}

\thy{ This new configuration represents a +5\% measurable improvement over the failed ``strongest'' configuration (Config A, 55\%). The expert used a multi-step diagnostic loop: he (1) diagnosed the \texttt{emb-8B} model's ``Paradox of Riches'' failure pattern, (2) formed a new ``less-noisy'' idea, and (3) confirmed it by finding a final configuration that is measurably better, cheaper, and more precise. }

\section{Expert Evaluation}
We conducted a qualitative study with domain experts to evaluate our system. Specifically, we aimed to assess the effectiveness of its individual views for analysis and the usability of the overall workflow.

\subsection{Participants}
We invited four domain experts (E3-E6) for this evaluation. E3 and E4 were the two industry practitioners from our formative study (\cref{sec:formativeStudy}). The other two experts, E5 and E6, were new participants from industry. They specialize in AI agent development and are experienced with RAG technologies.

\subsection{Procedure and Tasks}
The evaluation with each expert was structured into three phases. The session began with a 15-minute introduction, during which we presented the system's background and core motivation. We also provided a live demonstration of each view. After this, participants entered a 30-minute task-oriented exploration phase. We guided them with the two case studies presented in \cref{sec:case study}. While following the main path, participants were encouraged to explore freely and vocalize their reasoning and any discoveries with a think-aloud protocol. The entire interactive session was audio- and screen-recorded for subsequent analysis. The session concluded with a 20-minute questionnaire and a semi-structured interview. Participants rated the system on a five-point Likert scale, assessing the effectiveness of the views, the system's usability, and the coherence of the overall workflow. The final interview allowed us to gather more in-depth qualitative feedback and elaborate on their questionnaire responses.

\subsection{Results Analysis}
\begin{figure}[t]
  \centering
  \includegraphics[width=1.0\columnwidth]{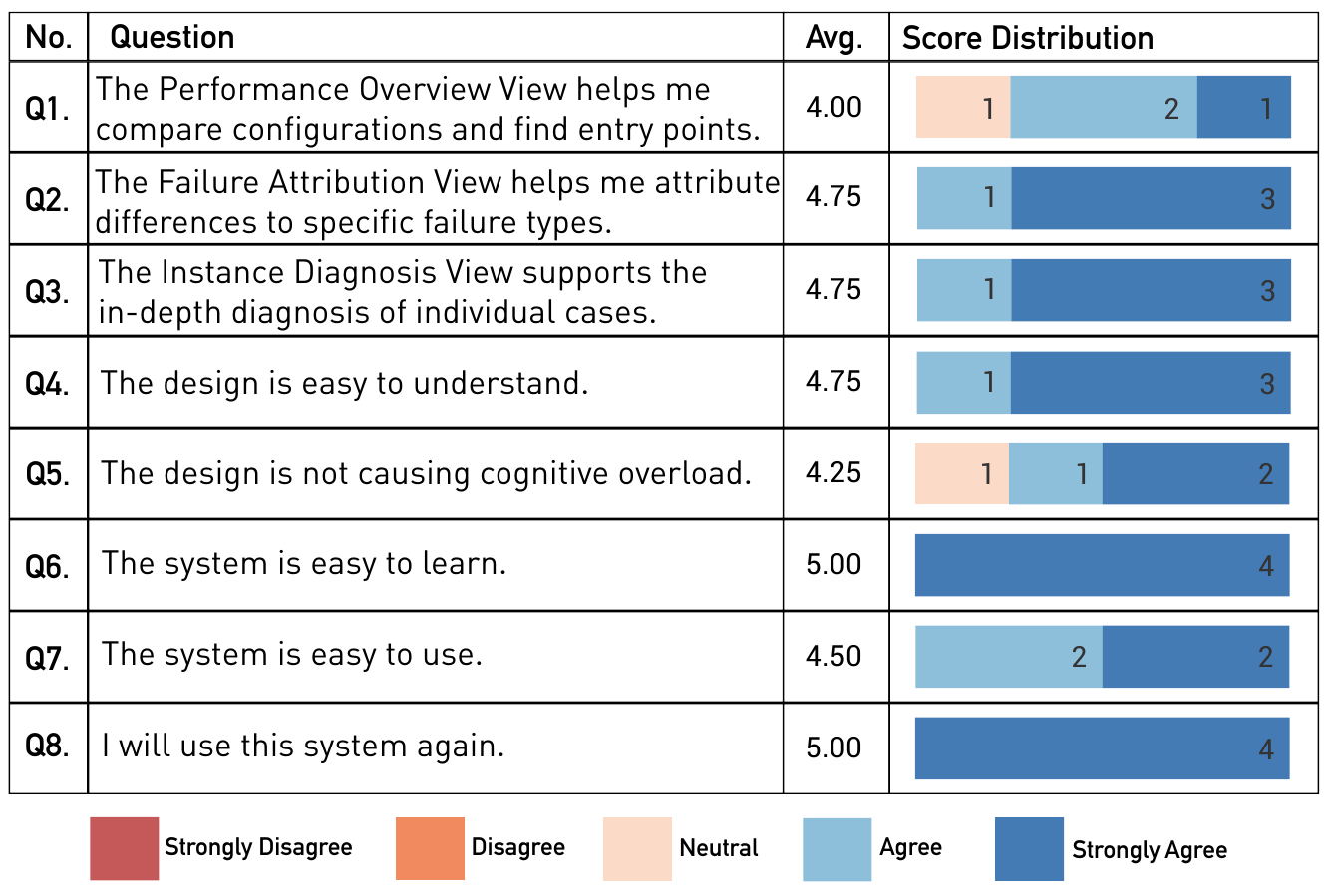}
  \caption{The results of the expert interview questionnaire regarding the effectiveness and usability of the RAGExplorer and workflow.}
  \vspace{-4pt}
  \label{fig:result}
\end{figure}

This section analyzes the expert feedback and questionnaire data (\cref{fig:result}). We evaluate the effectiveness of each system view and discuss the overall usability and workflow.

\subsubsection{Effectiveness of System Views}
In this section, we detail the expert feedback on the three core views of our system. The evaluation reveals that while each view was considered valuable for its specific analytical purpose, experts also identified distinct usability challenges and offered targeted suggestions for improvement related to information density, terminological clarity, and workflow continuity.

\vspace{4pt}
\noindent
The \textbf{Performance Overview View} serves as a valuable entry point for analysis but imposes a visual burden (\textbf{Q1}, $\mu$=4.0/5). While E3 described it as a ``\textit{perfect start}'' for finding patterns and generating hypotheses, E4 felt ``\textit{lost}'' among the numerous configuration points. E5 mentioned ``\textit{it's hard to quickly locate and compare the configurations I want to control with so many points}''. Similarly, E6 noted ``\textit{finding specific configurations was difficult.}'' E5 suggested adding direct filtering controls like a drop-down box, while E6 proposed ``\textit{enhancing the visual encoding to make performance differences more apparent}''.

\vspace{4pt}
\noindent
The \textbf{Failure Attribution View} was highly rated for its ability to identify differences between configurations (\textbf{Q2}, $\mu$=4.75/5). E3 highlighted its ability to help him ``\textit{lock down}'' key performance differences and praised the fluid ``\textit{brushing and linking}'' interaction. E5 noted that ``\textit{the flows within the Sankey diagram allowed me to very quickly see changes}''. But E4 was ``\textit{confused}'' by the specialized definitions of the Failure Points. He explained that a RAG expert is not necessarily an expert in failure attribution and may struggle to understand by its name alone. E3 suggested an insight module to generate textual summaries, and E6 built on this by proposing a ``\textit{one-click analysis}'' feature to automatically highlight the most significant changes.

\vspace{4pt}
\noindent
The \textbf{Instance Diagnosis View} was also confirmed for its utility in in-depth case validation (\textbf{Q3}, $\mu$=4.75/5). E5 praised the dual-axis design, stating it answered a key diagnostic question ``\textit{at a glance}'': ``\textit{whether a document was retrieved but failed due to low ranking}''. Conversely, E3 found the experience ``\textit{abrupt}'' when a specific case contradicted a hypothesis proposed previously. This situation forced him to check numerous cases manually, to determine ``\textit{if a contradictory case was an outlier or a general pattern}''. Experts suggested an information hierarchy with summaries (E3) and collapsible text regions (E4). Additionally, E3 recommended using an LLM to explain ``\textit{how this evidence contributes to the conclusion}'' rather than simply highlighting text. E4 also proposed a feature to track a document chunk's rank change.

\subsubsection{Overall Workflow and System Usability}
Beyond the individual views, we assessed the user experience, focusing on the system's overall workflow and usability. Experts universally praised the system's learnability and expressed a strong willingness to reuse it. However, this positive reception was balanced by a nuanced discussion of cognitive load and usability, which received more varied ratings and highlighted a tension between the system's analytical power and the mental effort it requires.

\vspace{4pt}
\noindent
\textbf{Design Clarity and Cognitive Load.}
Experts found the system easy to understand (\textbf{Q4}, $\mu$=4.75/5). E3 described ``\textit{the layout as clear but noted a risk of information overload in the Text Details Panel.}'' Cognitive load received a more varied rating (\textbf{Q5}, $\mu$=4.25/5). E4 explained that the system requires him to ``\textit{actively think and find the reason}'' for observed phenomena. E5 felt the pressure came ``\textit{specifically from the information density of the Performance Overview View}''. In contrast, E6 finds the information level appropriate and not overwhelming.

\vspace{4pt}
\noindent
\textbf{Learnability and Usability.}
The system's learnability received a perfect score from all experts (\textbf{Q6}, $\mu$=5.0/5). E6 commented that he could ``\textit{tell what it does at a glance}''. Usability, in contrast, was rated more variably (\textbf{Q7}, $\mu$=4.5/5). E5 noted it was very effective for an expert user wanting to ``\textit{verify a conclusion I had already predicted.}'' And E6 called the interaction ``\textit{simple and direct}''. However, E3 noted that ``\textit{the pairwise-comparison design required repeated analyses}'' and ``\textit{imposed a heavy reading burden}''. E4 indicated insufficient system guidance in certain views.

\vspace{4pt}
\noindent
\textbf{Overall Value and Willingness to Reuse.}
All experts are willing to reuse (\textbf{Q8}, $\mu$=5.0/5). E3 called it a ``\textit{huge improvement}'' over traditional analysis, noting it can accelerate the process and provide ``\textit{convincing visual evidence}'' for team collaboration. E4 considered it a valuable ``\textit{guided tuning tool}'' for scenario-specific optimization, since no ``\textit{silver bullet}'' for RAG exists. E5 described it as a practical ``\textit{debug tool for RAG workflow}''. E6 added that ``\textit{the system's comparative nature is ideal for facilitating analysis akin to ablation studies.}'' Finally, E3 suggested that ``\textit{the system's analytical paradigm has good extensibility for more complex scenarios, such as GraphRAG}''.

\subsection{Findings}
From the expert evaluation and case studies, we derive four key findings. These insights operate on two distinct levels: the first two reveal fundamental principles about the RAG optimization process, discovered through our system. The latter two are synthesized from the experts' feedback on the analytical workflow, identifying a valuable new paradigm for AI diagnostics and pointing toward a future of LLM-augmented collaborative analysis.

\vspace{4pt}
\noindent
\thy{
\textbf{Aggregate Statistics Mask Hidden Performance Gains}
A key finding from our case study is that aggregate performance metrics (like \texttt{Accuracy}, \texttt{MAP}, or \texttt{MRR}) are insufficient and can be highly misleading for RAG optimization. In our case, standard ``simple statistics'' highlighted an apparent lack of difference between several configurations, showing nearly identical scores that suggested a key parameter was irrelevant. The Failure Attribution View, however, provided the true causal explanation. It revealed a hidden failure mode transfer: one configuration was proven to be objectively superior in recall (fewer \texttt{FP2} errors), but this gain was being offset by a new bottleneck in the reranking stage (more \texttt{FP3} errors). This reveals that aggregate metrics can mask objective performance gains in early pipeline stages. These gains are often simply canceled out by a new, downstream bottleneck, a performance pattern our system is designed to reveal.
}

\vspace{4pt}
\noindent
\thy{
\textbf{Component Synergy Outweighs Individual Strength.}
Our analysis also revealed that simply combining the ``strongest'' components does not guarantee the best performance. RAGExplorer's Performance Overview first highlighted this counter-intuitive finding, where a ``strongest'' configuration (\texttt{Qwen3-Embedding-8B} + \texttt{Qwen3-Reranker-8B} + \texttt{chunk size=2000}) was measurably outperformed by a ``Budget'' one (\texttt{Qwen3-Embedding-0.6B} + \texttt{Qwen3-Reranker-0.6B} + \texttt{chunk size=2000}).
This finding offers a significant practical insight: RAG optimization is not about maximizing individual component strength, but about finding optimal synergy. A ``weaker'' component that provides a ``cleaner'' signal can be a more resource-effective and higher-performing strategy than simply deploying the largest models.
}

\vspace{4pt}
\noindent
\textbf{The Value of a Comparative and Hierarchical Workflow.}
Our study reveals the effectiveness of an analytical workflow that integrates pairwise comparison with hierarchical analysis for diagnosing complex RAG systems. E6 aptly characterized this workflow's value as enabling ``\textit{ablation studies}''. This approach allows developers to trace the systemic impact of a single component change, from a high-level performance shift down to its root causes within the pipeline. This integrated paradigm moves beyond the limitations of analyzing metrics in isolation, providing a causal, step-by-step narrative of how a configuration choice influences specific failure modes. The finding confirms that for multi-component AI systems, such a workflow is critical for moving from simple performance comparison to deep, actionable diagnostics.

\vspace{4pt}
\noindent
\textbf{The Need for LLM-Augmented Collaborative Analysis.}
A consistent theme from expert feedback was the demand for the system to evolve from a passive visualization tool into a proactive analytical partner. This signals an emerging need for a new paradigm of human-AI collaborative analysis where the exploratory burden is shared. Experts proposed features for automated discovery, such as an ``\textit{insight module}'' (E3) and a ``\textit{one-click analysis}'' (E6), to automatically surface the most significant performance changes. More profoundly, feedback called for deeper explanatory augmentation. A key suggestion was to leverage an LLM to explain ``\textit{how this evidence contributes to the conclusion}'' (E3), helping users interpret complex or contradictory cases. These suggestions collectively point toward a future where interactive visualization is augmented by LLM-driven reasoning, allowing human experts to concentrate on strategic decision-making and validation.

\section{Discussion}
The analysis of RAG systems is complicated by the modularity of their architecture, where end-to-end performance is an emergent property of interacting components. 
This work introduces RAGExplorer, a visual analytics system designed to address this complexity.
We contribute a methodology for comparative diagnosis, which shifts the analytical objective from single-configuration debugging to a systematic analysis of performance differences and failure mode transitions between multiple configurations.
This approach provides a structured paradigm to explore the design space and understand its inherent trade-offs. 


\vspace{4pt}
\noindent
\textbf{Generalizability.}
\wz{In this work, we have demonstrated how our system effectively works on a fixed set of RAG configurations. 
To support generalization with more user-defined configurations, the RAGExplorer is developed as a plug-in system where users can register custom components to extend additional configurations.
 On this basis, our visual analytics paradigm can be further generalized to other modular RAG pipelines.} 
 Its comparative visualization is effective for diagnosing system-level behaviors in multi-stage pipelines, e.g., agentic workflows.
 By focusing on visualizing the differences between configurations, such as failure transitions or content disparities, our approach allows analysts to debug emergent issues that are invisible to component-level inspection. 
 Benefiting from our visual and interactive interface, this comparative analysis can be further strengthened by enabling active intervention for users to directly test causal hypotheses.

 
\vspace{4pt}
\noindent
\textbf{Applicability.}
\wz{Our framework relies on the ground-truth data to calculate the evaluation metrics and failure points.
To improve applicability in production where the ground truth is missing, RAGExplorer can be adapted to operate using LLM-as-a-Judge evaluators for metrics like context relevance and faithfulness. 
This preserves the core capability of understanding the impact of configuration changes for comparative diagnosis by visualizing shifts between these automatically assessed metrics, thus relaxing the dependency on annotated ground-truth data.}

\vspace{4pt}
\noindent
\textbf{Scalability.}
\wz{The current system is designed for the focused, in-depth comparison of a small number of configurations, which is typical for root-cause analysis.}
While the Performance Overview View supports the initial exploration of dozens of configurations, the approach does not scale directly to thousands of experiments common in large-scale hyperparameter searches. 
Addressing this would require future work on automated methods to supplement the visual exploration, such as algorithms to detect salient patterns across the configuration space or to recommend the most diagnostically insightful comparisons to the user.
\thy{Further future work could incorporate sampling\cite{DBLP:conf/kdd/ThorntonHHL13,DBLP:journals/corr/abs-2502-18635} or hierarchical search of the parameter space to reduce computational overhead.
}


\vspace{4pt}
\noindent
\textbf{Study Limitations.}
\wz{The evaluation was scoped to question-answering tasks with annotated ground-truth data. 
While we have discussed paths to generalize, the system's effectiveness on open-ended tasks without such annotations remains a key area for future work.} 
Besides, our user study was conducted exclusively with domain experts. 
The learnability and utility of RAGExplorer for novice users, who may lack deep domain knowledge, is an important area for future investigation.

\section{Conclusion}
In this paper, we introduced RAGExplorer, a visual analytics system designed to untangle the complexity of configuring RAG systems. 
We presented a novel analytical workflow that shifts the focus from debugging single pipelines to the systematic, comparative diagnosis of multiple configurations. 
This approach empowers developers to move seamlessly from a high-level performance overview to an in-depth, interactive investigation of instance-level failures.
Our case studies and user evaluation demonstrated that this comparative paradigm effectively uncovers critical, non-obvious insights into RAG system behavior. 
These results underscore the necessity of a holistic, system-level approach to optimization.
By providing a structured framework and an interactive interface for comparative analysis, RAGExplorer not only offers a practical tool for AI developers but also establishes a visual analytics paradigm poised to bring clarity to the emergent behaviors of a wide range of complex, modular AI systems.

\acknowledgments{
We thank anonymous reviewers for their insightful reviews. This work was supported by the National Key R\&D Program of China (Grant Nos. 2024YFB4505500 and 2024YFB4505503), the National Natural Science Foundation of China (Grant Nos. 62132017, 62421003, and 62302435), and the Zhejiang Provincial Natural Science Foundation of China (Grant No. LD24F020011).
}

\bibliographystyle{abbrv-doi-hyperref}

\bibliography{template}

\end{document}